\documentclass[aps,prl,twocolumn,superscriptaddress,showpacs]{revtex4}
 \usepackage{graphicx}
\usepackage{afterpage} \usepackage{dcolumn} \usepackage{bm}

\begin{document}

\preprint{APS/123-QED}

\title{Electron-phonon interaction and charge carrier mass
enhancement in SrTiO$_3$}

\author{J.L.M.~van~Mechelen}

\affiliation{ D\'epartement de Physique de la Mati\`ere
Condens\'ee, Universit\'e de Gen\`eve, Gen\`eve, Switzerland.}

\author{D.~van~der~Marel}

\affiliation{ D\'epartement de Physique de la Mati\`ere
Condens\'ee, Universit\'e de Gen\`eve, Gen\`eve, Switzerland.}

\author{C.~Grimaldi}

\affiliation{ D\'epartement de Physique de la Mati\`ere
Condens\'ee, Universit\'e de Gen\`eve, Gen\`eve, Switzerland.}

\affiliation{LPM, Ecole Polytechnique F\'ed\'erale de Lausanne,
Lausanne, Switzerland.}

\author{A.B.~Kuzmenko}

\affiliation{ D\'epartement de Physique de la Mati\`ere
Condens\'ee, Universit\'e de Gen\`eve, Gen\`eve, Switzerland.}

\author{N.P.~Armitage}

\affiliation{ D\'epartement de Physique de la Mati\`ere
Condens\'ee, Universit\'e de Gen\`eve, Gen\`eve, Switzerland.}

\affiliation{Department of Physics and Astronomy, The Johns
Hopkins University, Baltimore, USA.}

\author{N.~Reyren}

\affiliation{ D\'epartement de Physique de la Mati\`ere
Condens\'ee, Universit\'e de Gen\`eve, Gen\`eve, Switzerland.}

\author{H.~Hagemann}

\affiliation{D\'epartement de Chimie Physique, Universit\'e de
Gen\`eve, Gen\`eve, Switzerland.}

\author{I.I.~Mazin}

\affiliation{Center for Computational Materials Science, Naval
Research Laboratory, Washington D.C., USA.}


\pacs{71.38.-k, 72.20.-i, 78.20.-e}

\begin{abstract}
We report a comprehensive THz, infrared and optical study of Nb doped SrTiO$_3$ as well as DC conductivity and Hall effect measurements. Our THz spectra at 7~K show the presence of a very narrow ($<2$ meV) Drude peak, the spectral weight of which shows approximately a factor of three enhancement of the band mass  for all carrier concentrations. The missing spectral weight is regained in a broad `mid-infrared' band which originates from electron-phonon coupling. We find no evidence of a particularly large electron-phonon coupling that would result in small polaron formation. Analysis of the results yields an electron-phonon coupling parameter of an intermediate strength, $\alpha \approx 4$.
\end{abstract}

\maketitle

Electron-phonon coupling in the perovskites is a subject of much recent interest due to the controversy over its relevance in the phenomena of multiferroicity, ferroelectricity, superconductivity and colossal magnetoresistance~\cite{lanzara,alexandrov,hwang,manella,bednorz&muller}. Despite much progress, full understanding of the physics of electron-phonon coupling in perovskites is still lacking because of additional crystallographic complexities of many materials involved (breathing, tilting and rotational distortions, ferroelectric symmetry breaking), magnetism, complex electronic effects (strong correlations), and also because of the lack of high-accuracy spectroscopic measurements specifically designed to probe electron-phonon coupling.

With this in mind,  we have studied a prototypical perovskite oxide, SrTi$_{1-x}$Nb$_{x}$O$_{3}$ with $0\leq x\leq 0.02$. SrTiO$_3$ is an insulator ($\Delta=3.25$ eV) with the conduction band formed by the Ti 3d states. These are  split by the crystal field so that the three t$_{2g}$ states become occupied when the material is electron-doped by substituting pentavalent Nb for tetravalent Ti. For $0.0005\leq x\leq0.02$ SrTi$_{1-x}$Nb$_{x}$O$_{3}$ becomes  superconducting at a $T_c$ of typically $\sim0.3$ K~\cite{koonce,note1}, and at most 1.2 K~\cite{bednorz&muller}. Characterized by the threefold degeneracy of the conduction bands and the high lattice polarizability, electron doped SrTiO$_3$ provides a perfect opportunity for the study of electron-phonon coupling and polaron formation in an archetypal perovskite~\cite{calvani01,eagles}.

\begin{figure}[t]
\centering
\includegraphics[width=\linewidth]{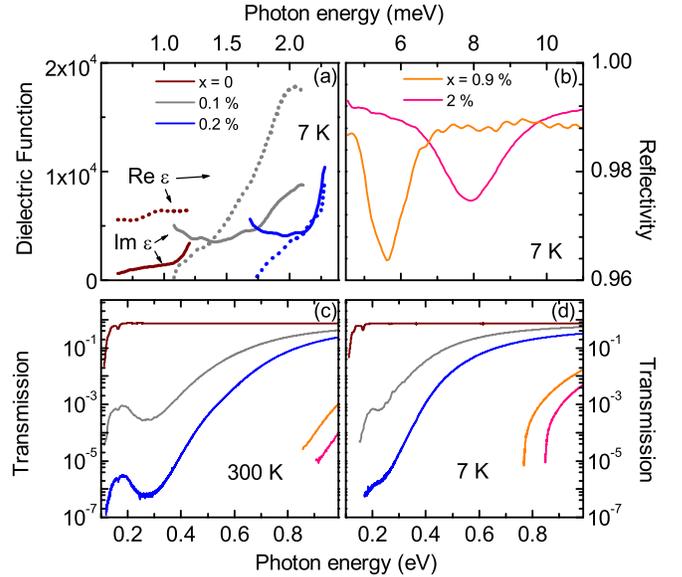}
\caption{Selected the experimental data. A
very narrow Drude peak at 7K can be seen in both (a),
$\epsilon_1(\omega)$ (dotted curves) and $\epsilon_2(\omega)$
(solid curves) in the THz range for $x=0.1\%$ and 0.2\%, and
 (b), the upturn in the reflectivity at low energies for $x=0.9\%$
and 2.0\%. (c-d), Transmission coefficient at (c), 300 K
and (d), 7 K showing the absorption in the mid-infrared.}
\label{fig_rawdata}
\end{figure}

One of the fingerprints of an ultrastrong electron-phonon coupling is the formation of small polarons that are observable in the form of a gigantic electron mass renormalization. The renormalized mass can be obtained by measuring the optical spectral weight of the Drude peak, which in proper units is equal to $n/m^*$. Besides renormalizing the `coherent' (Drude) part of the spectrum,  the electron-phonon coupling is responsible for the `incoherent' contributions at higher energies, resulting in a multiphonon absorption band in the mid-infared.

To address these issues, we have measured
the optical reflectivity and transmission (Fig. \ref{fig_rawdata})
of double side polished, $5\times5$ mm$^2$ ((100) face) single
crystals of Sr$_{1-x}$Nb$_{x}$TiO$_3$ between
300 K and 7 K by time-domain THz spectroscopy (TPI spectra 1000,
TeraView Ltd.), Fourier transform infrared spectroscopy and
photometric IR-UV spectroscopy, between 0.3 meV and 7 eV. 
To obtain a detectable transmission we have used for each
composition several samples of different thickness (8 --
60 $\mu$m), adopted to the spectral range and value of the optical
transmission. The Hall carrier concentrations were $0.105$\%, $0.196$\%, $0.875$\% and $2.00$\% at 7 K, which were within 5 $\%$ of those measured by the  wavelength-dispersive X-ray spectroscopy, and within 12 $\%$ of the Nb concentration specified by the supplier (0.1\%, 0.2\%, 1.0\% and 2.0\%, respectively, Crystec, Berlin). Compared to earlier measurements~\cite{barker,gervais,crandles,calvani93,bi}, we expand the spectral range further into the far-infrared and THz band. Fig. \ref{fig_rawdata}a shows both components of the dielectric function, $\epsilon_1(\omega)$ and $\epsilon_2(\omega)$, in the THz range at 7 K, as obtained by a direct inversion of the measured transmission amplitude and phase. From the DC resistivity, the reflectivity and the transmission data we calculated the real and imaginary part of the optical conductivity using a generalized Kramers-Kronig routine~\cite{kuzmenko05}.

\begin{figure}[t]

\centering
\includegraphics[width=7.5cm]{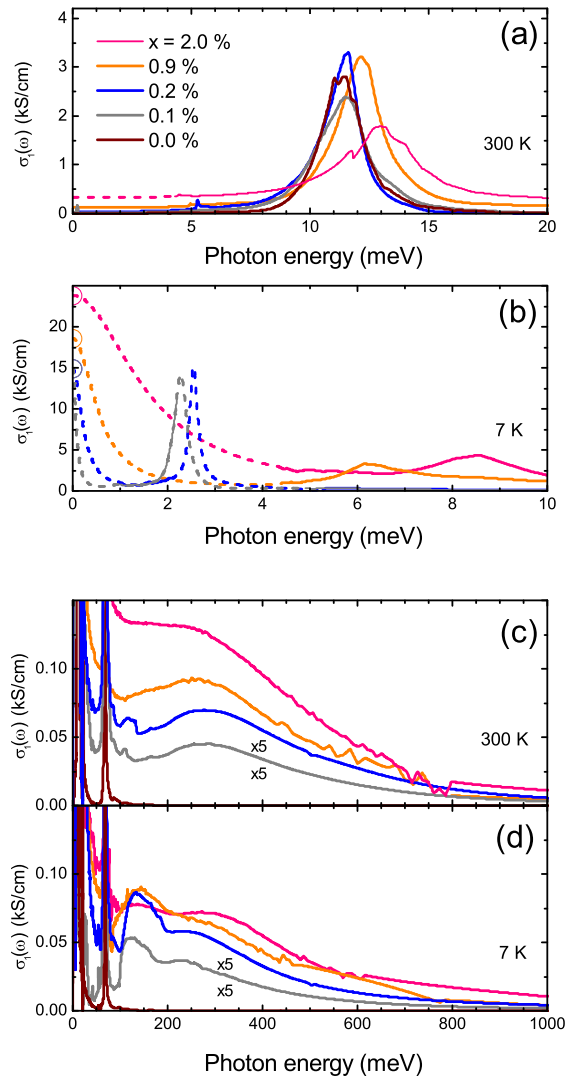}
\caption{The optical conductivity of
SrTi$_{1-x}$Nb$_{x}$O$_3$ for $x=$ 0, 0.1\%, 0.2\%, 1\% and 2\% at
300 and 7 K. The dashed parts of the curves in (a-b) are
interpolations fitted to the DC resistivity, reflectivity,
transmission and $\epsilon_1(\omega)+i\epsilon_2(\omega)$ obtained
from the THz transmission amplitude and phase. Open symbols are
the experimental DC conductivities. (c-d). For clarity, the
mid-infrared conductivities of $x = 0.1\%$ and 0.2\% are magnified
by a factor 5.}
\label{fig_sigma}
\end{figure}

The only subgap contributions to the optical conductivity of undoped SrTiO$_3$ (Fig. \ref{fig_sigma}) are the three infrared active phonons at 11.0, 21.8 and 67.6 meV (at room temperature). The lowest one exhibits a strong red shift upon cooling, and saturates at about 2.3 meV at 7 K. Upon doping this mode hardens for all temperatures, reaching 8.5 meV at 7 K at 2$\%$ doping. For the present discussion, the most important effect of substituting Nb is doping electrons, which yields a clearly distinguishable Drude peak that is broad at 300 K but gets extremely narrow at low temperatures. For the lowest dopings, this can be directly seen from the upturn of $\epsilon_2(\omega)$ for $\hbar\omega<2$ meV (Fig. \ref{fig_rawdata}a). For the 0.9\% and 2\% doped samples, the metallic conductivity is  directly visible from the increase of the reflectivity below the soft optical phonon (Fig.~\ref{fig_rawdata}b). The third effect of doping is the occurrence of a broad asymmetric mid-infrared absorption that increases with doping. At 7 K, a sharper structure shows up on the low-energy side of this band (Fig. \ref{fig_sigma}c-d). Using first principles LDA calculations~\cite{LDA} we have verified that interband transitions within the $t_{2g}$ manifold are orders of magnitude too weak to account for the observed intensity. We show below that the observed lineshape can be understood as a multiphonon sideband of the free carrier (Drude) response.
\begin{figure}[t]
\centering \includegraphics[width=\linewidth]{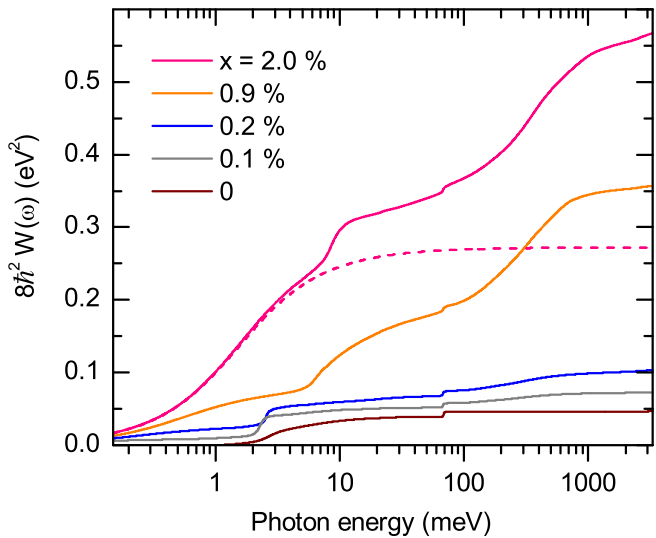}
\caption{The integrated optical spectral weight of
SrTi$_{1-x}$Nb$_{x}$O$_3$ as a function of photon energy and
relative to the charge carrier concentration, at 7 K. The Drude
spectral weight is separately shown for $x=2\%$ by the dotted line. }
\label{fig_sw}
\end{figure}

As mentioned before, the effective mass of the charge carriers
$m^*$ can be obtained by analyzing the Drude spectral weight.
In Fig. \ref{fig_sw} we show the spectral weight $W(\omega)$,
defined as 
\begin{equation}\label{Fsumrule}
     W(\omega_c)\equiv \int_0^{\omega_c}\sigma_1(\omega)d\omega.
\end{equation}
Neglecting the optical transitions from the core, the electronic contribution $8W(\infty)= \omega_p^2=4\pi n e^2/m$, where $e$ and $m$ are the free electron charge and mass, and $n$ is the valence electron density. The contribution of the infrared phonons manifests itself in $W(\omega)$ as steps, {\em e.g.} at 8.5 meV in the $2\%$ doped sample (Fig. \ref{fig_sw}). Such a strong phonon close to the Drude peak makes a decomposition of their respective spectral weights challenging, and requires experimental data that extend beyond the soft phonon frequency unlike previous experimental data. Here, (\textit{i}) direct measurements of $\epsilon_1(\omega)$ and $\epsilon_2(\omega)$ below 3 meV by time-domain THz spectroscopy (for the $x=0.1\%$ and 0.2$\%$ samples) and (\textit{ii}) far-infrared reflectivity between 4 and 10 meV (for the $x=0.9\%$ and 2$\%$ samples) shown in Fig. \ref{fig_rawdata} allowed us to extract the Drude spectral weight using several methods.

\begin{figure}[t]
\centering \includegraphics[width=\linewidth]{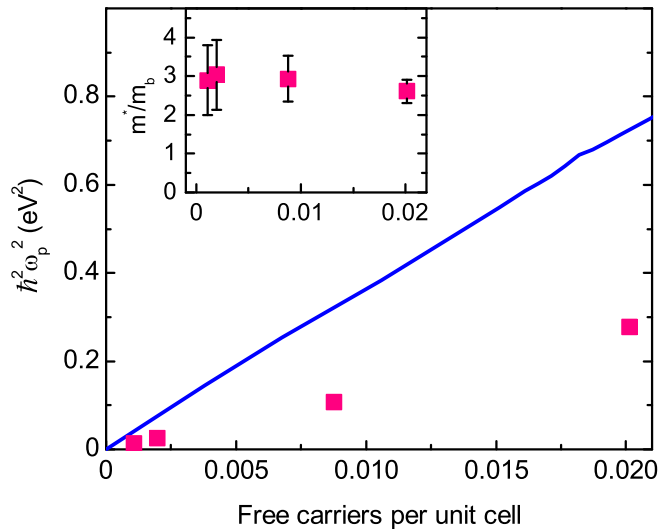}
\caption{Experimental plasma frequency at 7 K derived from the spectral
weight $W(\omega)$ of the Drude peak (squares) as a function of
the free carrier concentration (from Hall effect) together with
that from the band structure calculations~\cite{LDA}
(solid line). The inset shows the
corresponding effective carrier mass $m^*/m_b$.}
\label{fig_mass}
\end{figure}

For the two lowest dopings we apply the partial sum rule (Eq.\ref{Fsumrule}) with cutoff frequencies $\hbar\omega_c=1.5$ meV (0.1\%) and 2.0 meV (0.2\%), which are larger than the Drude relaxation rate but below the soft phonon (see Fig. 2b)~\cite{note2}, yielding $\hbar\omega_p=120\ \pm\ 20$ meV and $201\ \pm\ 90$ meV, respectively. These values slightly change to $\hbar\omega_p=116\ \pm\ 18$ meV and $159\ \pm\ 23$ meV, respectively, when we carry out a simultaneous Drude-Lorentz fit to these data together with reflectivity and transmission data at higher frequencies and the DC resistivity. For the two higher
dopings, no samples transparent in the THz band could be prepared,
but the soft-phonon frequency shifts with doping into a range where reliable infrared
reflectivity can be obtained, resulting in the dips in reflectivity seen in Fig.
1b. Thus, a standard Kramers-Kroning analysis of
the reflectivity can be used as well as Drude-Lorentz fitting. 
Both methods give essentially identical values, $\hbar\omega_p=327\ \pm\ 67$ meV
and $527\ \pm\ 61$ meV for 0.9$\%$ and 2$\%$ doping, respectively, shown by the red squares in Fig. 4. Defining the mass renormalization of the
charge carriers as the ratio of the LDA and the experimental values of the
 squared plasma-frequency,
$m^*/m_b\equiv\omega_{p,\text{LDA}}^2/\omega_{p,\text{exp}}^2$, we
observe a threefold mass enhancement, {\em i.e.},
$m^*/m_b = 2.9 \pm 0.3$ for the low charge carrier densities
considered in the present study (Fig. \ref{fig_mass}, inset).

For  $0.001\leq x\leq 0.02$, the band-structure value~\cite{note3} of the Fermi energy, $E_F$, ranges from 0.02 to 0.1 eV, relative to the bottom of the lowest $t_{2g}$ band, which coincides with the energy of the relevant longitudinal optical phonon, $\hbar\omega_{\text{LO}}$ = 0.1 eV~\cite{servoin,verbist}. This is a difficult parameter range to
describe theoretically where none of the usual approximations,
{\em i.e.}, $\hbar\omega_{\text{LO}}/E_F \ll 1$ nor
$\hbar\omega_{\text{LO}}/E_F \gg 1$ are applicable.  For the
present discussion we consider the latter limit, appropriate for
the lower doped samples. It takes as a starting point the model of
a single polaron \cite{feynman,mahan}, characterized by the
coupling constant $\alpha$. Within a variational approximation,
Feynman~\cite{feynman} deduced the relation $m^*(\alpha)$ valid in
the range $0< \alpha < 12$. Applying Feynman's relation to the
experimental $m^*$ values, we obtain a doping independent value of
$\alpha\approx 4$. This is considerably larger than the value
expected for a purely Fr\"ohlich-type interaction, for which (using experimental optical parameters of SrTiO$_3$)
$\alpha = \sqrt{m_b/m_e} \sqrt{Ry/\omega_{\text{LO}}}
(\epsilon_{\infty}^{-1}-\epsilon_{s}^{-1})= 2.1$.

The optical conductivity shows a broad absorption band between 0.1 and 1 eV (Fig. 2c and 2d). The intensity in this mid-infrared band compensates the deficit of Drude spectral weight and the f-sum rule, $8W(\infty)=\omega_p^2$, is almost satisfied at 3 eV, as can be seen from Fig.~\ref{fig_sw}~\cite{note4}. Such a spectral weight redistribution between the `coherent' (Drude) and the `incoherent' (mid-infrared) contributions of the optical conductivity suggests that the mass enhancement reflects coupling of electrons to bosonic degrees of freedom, most probably phonons. A spectral shape similar to the observed mid-infrared band at 300 K (Fig. 2c) was calculated by Emin~\cite{emin}, who considered the photo-ionization spectrum out of the potential well formed by the lattice deformation of the polaron. Upon cooling, the maximum of this band remains between 200 and 350 meV, and at 7K an additional peak emerges on the low-energy side with a maximum varying from 120 (0.1$\%$) to 140 meV (2$\%$). Devreese {\em et al.}~\cite{kartheuser69,devreese72,iadonisi84} have numerically found a broad mid-infrared band with (for intermediate couplings) a peak at lower energies. They explain it as a multiphonon band and the peak as a relaxed state of an electron optically excited within the polaronic potential well. The model predicts the maxima of the two bands to be at $0.065\alpha^2\hbar\omega_{\text{LO}}$ and $0.14\alpha^2\hbar\omega_{\text{LO}}$. With the value $\alpha\approx 4$ that we deduced above this corresponds to $110$ meV and $230$ meV, consistent with the experimentally observed positions (Fig. 2d).

We conclude that the electron-phonon coupling leads to a threefold mass enhancement in electron doped SrTiO$_3$ for charge carrier concentrations between 0.1\% and 2\% per unit cell. The `missing' Drude spectral weight, as measured directly by THz time-domain transmission spectroscopy, is recovered in the mid-infrared side-bands resulting from the electron-phonon coupling interaction. Analysis of the spectral weight transfer yields a polaronic electron-phonon coupling parameter of intermediate strength, $\alpha \approx 4$. This suggests that the charge transport in electron doped SrTiO$_3$ could be carried by large polarons. We find no evidence for an extremely strong electron-phonon coupling  or small polarons in the considered doping range. Our observations may have important implications for the electron-phonon coupling in perovskites in general, including ferroelectrics, multiferroics, colossal magnetoresistance materials and high-$T_c$ cuprates.

This work was supported by the Swiss National Science Foundation through the NCCR `Materials with Novel Electronic Properties' (MaNEP). NPA has been supported via the NSF's International Research Fellows program. We gratefully acknowledge stimulating discussions with J.-M.~Triscone, K.S.~Takahashi, T.~Giamarchi, J.T.~Devreese, J.~Lorenzana and \O.~Fischer.


\begin{thebibliography}{99}
\bibitem{note4} In order to account for the lattice vibrational contributions (assumed to be doping independent) to the sum rule at $\omega_c=3$ eV we subtracted, for all dopings, the corresponding value of undoped SrTiO$_3$.
\end{thebibliography}
\end{document}